\documentclass[aps,prd,twocolumn,showpacs]{revtex4}
\usepackage{ulem}
\usepackage{color}
\usepackage{graphicx}
\usepackage{epsfig}
\usepackage{bm}
\def\nn {\nonumber}

\newcommand{\be}{\begin{equation}}
\newcommand{\ee}{\end{equation}}
\newcommand{\bea}{\begin{eqnarray}}
\newcommand{\eea}{\end{eqnarray}}

\newcommand{\ep}{\epsilon}

\newcommand{\om}{\omega}

\newcommand{\ov}{\overline}

\newcommand{\ds}{\partial \!\!\! /}

\newcommand{\ks}{k \!\!\! /}

\newcommand{\bk}{\boldsymbol{k}}
\newcommand{\bq}{\boldsymbol{q}}
\newcommand{\bl}{\boldsymbol{l}}
\newcommand{\bp}{\boldsymbol{p}}
\newcommand{\bu}{\boldsymbol{u}}
\newcommand{\bpi}{\boldsymbol{\pi}}
\newcommand{\btau}{\boldsymbol{\tau}}

\newcommand{\del}{\partial}

\begin{document}

\title{Transport responses from rate of decay and scattering processes 
in the Nambu--Jona-Lasinio model}
\author{Sabyasachi Ghosh$^1$, Fernando E. Serna$^{2,3}$, Aman Abhishek$^{4,5}$, 
Gast\~ao Krein$^2$, Hiranmaya Mishra$^4$}
\affiliation{$^1$ Indian Institute of Technology Bhilai, GEC Campus, Sejbahar, Raipur 492015, 
Chhattisgarh, India}
\affiliation{$^2$Instituto de Fisica Teorica, Universidade Estadual Paulista, 
Rua Dr. Bento Teobaldo Ferraz, 271, 01140-070 S\~{a}o Paulo, SP, Brazil}
\affiliation{$^3$Instituto Tecnol\'{o}gico de Aeron\'{a}utica, DCTA,
12228-900 S\~{a}o Jos\'{e} dos Campos, SP, Brazil}
\affiliation{$^4$Theory Division, Physical Research Laboratory, 
Navrangpura, Ahmedabad 380 009, India}
\affiliation{$^5$Indian Institute of Technology, 
Gandhinagar-382355, Gujarat, India}

\begin{abstract}
We have calculated quark and anti-quark relaxation time by considering
different possible elastic and inelastic scatterings in the medium.
Comparative role of these elastic and inelastic scatterings on different
transport coefficients are explored. The quark-meson effective interaction 
Lagrangian density in the framework of Nambu--Jona-Lasinio model is used for 
calculating both type of scatterings. Owing to a kinetic threshold, 
inelastic scatterings can only exist beyond the Mott line in temperature and 
chemical potential plane, whereas elastic scatterings occur in the entire plane. 
Interestingly, the strength of inelastic scatterings near and above Mott line 
becomes so strong that medium behaves like a perfect fluid, in that
all transport coefficients become very small. 
\end{abstract}

\maketitle
\section{Introduction}

The study of transport coefficients at finite temperature and baryon 
density of strongly interacting matter is a subject 
of current interest in different contexts. At high baryon density and 
low temperature, transport coefficients are relevant for the study 
of an array of phenomena in compact stars~\cite{Buballa:2014jta,Baym:2017whm},
and at high temperatures and low densities, they are relevant in the
context of heavy-ion collisions~\cite{Bratkovskaya:2017ssq}, mainly 
in connection with the strongly-coupled nature of the quark-gluon plasma (QGP)
produced in those collisions~\cite{Shuryak:2014zxa}. A weakly interacting QGP 
was a natural expectation at high temperatures since the early days of QCD
because of its asymptotic freedom property. However, that expectation seems
not to be realized at current collider energies, as the data on such collisions 
performed at the Relativistic Heavy Ion Collider (RHIC) and the Large Hadron Collider (LHC)
can be described in a first approximation by low-viscosity hydrodynamics, a
feature that implies a strongly-coupled medium. Here, the most relevant transport coefficient 
is shear viscosity, $\eta$. Still in the context of heavy-ion collisions,
another important transport coefficient is electrical conductivity, 
$\sigma$. This coefficient is relevant e.g. for describing the low mass dimuon 
enhancement~\cite{Rapp_rev,G_rev} measured by the NA60 collaboration~\cite{SPS} at CERN, 
a subject related to chiral symmetry restoration. Similar to electrical conductivity, 
thermal conductivity, $\kappa$, is still another relevant  transport 
coefficient. The effect of thermal conductivity on the hydrodynamical evolution 
of the QGP can not be neglected in situations of high baryon density, as 
thermal conduction takes place in the medium.

Currently, transport coefficients of the QGP can not be calculated directly from QCD
using analytical methods. Lattice QCD is delivering its first results on these quantities. 
For example, results for the electrical conductivity 
have been reported in Refs~\cite{LQCD_Buividovich,LQCD_Amato,LQCD_Burnier,
LQCD_Arts_2007,LQCD_Barndt,LQCD_Ding,LQCD_Gupta}, but the results still contain  
large uncertainties. Extraction of these coefficients from lattice simulations are very 
challenging  because because they are Minkowski-space dynamical quantities and the simulations 
are performed in Euclidean space{\textemdash}Ref.~\cite{Ratti:2018ksb} is 
a recent review on the subject. Effective models of QCD have been extensively used in the 
recent years and have been a fundamental source of a great deal of insight 
on these coefficients. By far, shear viscosity $\eta$ has attracted most of the
attention because of its central role in signalling the strongly coupled nature of 
the QGP{\textemdash}the literature is too extensive to be reviewed here and but a selective 
list of references, closely related to the present work, can be found 
in Refs.~\cite{{Nicola},{Marty},{Gavin},{Prakash},{Deb:2016myz},{Ghosh:2015mda},{G_CAPSS},{Weise2},{LKW},{klevansky},{klevansky2},{Redlich_NPA},{Abhishek:2017pkp},{Harutyunyan:2017ttz},{HMPQM1}}. 
Regarding $\sigma$, recent calculations find contradictory results for hadronic matter
(pion gas): while Ref.~\cite{Lee} finds the ratio $\sigma/T$ increasing with 
$T$, Refs.~\cite{Nicola_PRD,Nicola,G_El} find it decreasing with $T$. Calculations 
using models intended to describe simultaneously hadronic and partonic matter, like the 
Parton-Hadron-String-Dynamics model~\cite{Cassing} and the Nambu-Jona-Lasino (NJL)
model~\cite{Marty} predict $\sigma(T)/T$ decreasing with $T$ in hadronic phase 
and increasing in the partonic phase. A calculation employing a holographic 
model~\cite{Finazzo} and others employing transport simulations~\cite{Puglisi,Greif} find 
$\sigma(T)/T$ decreasing in the partonic phase. Regarding thermal conductivity, 
its has been addressed in a few references using different models~\cite{{Gavin},{Sarkar},{Nicola},{Prakash},
{Davesne},{Nam},{NJL_Iwasaki},{Deb:2016myz},{Marty},{CFL1},{CFL2}}. 

Dissipation is a dynamical effect that plays an important in defining the properties of 
transport coefficients. In a quasi-particle description of the strongly interacting medium, 
the main source of dissipation are the microscopic elastic and inelastic scattering processes 
involving the quasi-particles. In most of the models employed to date, the quasi-particles are
massive constituent quarks and scalar-isoscalar ($\sigma$) and pseudoscalar-isovector 
($\pi$) mesons, the degrees of freedom associated with the dynamical breaking of chiral symmetry. 
Earlier~\cite{klevansky,klevansky2} and more recent~\cite{Marty,Deb:2016myz,{Abhishek:2017pkp}} 
works have investigated the effects of $2\leftrightarrow 2$ type of elastic quark-quark scatterings. 
More recently, Refs.~\cite{Ghosh:2015mda,G_CAPSS,Weise2,LKW,Harutyunyan:2017ttz,HMPQM1} have investigated
the contributions of inelastic quak-meson processes of the type $1 \leftrightarrow 2$, finding
that those processes play an important role within a narrow temperature window near but above 
the Mott temperature, a temperature beyond which the pionic bound state delocalizes into its 
constituents. However, to date no work has investigated the joint effect of the
$2\leftrightarrow 2$ and $1\leftrightarrow 2$ types of scatterings. The present work provides
such an investigation within the dynamical framework of the Nambu--Jona-Lasinio (NJL) 
model.
This model allows us, in particular, to calculate
the $2\leftrightarrow 2$ and $1\leftrightarrow 2$ processes within the same underlying model
and, in addition, to take into account in a self-consistent manner the dynamics of chiral symmetry 
restoration as a function of temperature and baryon density. We find, in particular, that the 
self-consistency has a dramatic effect on the transport coefficients.

The paper is organized as follows. In the next section, we address the formalism part,
where we define the model and present the standard expressions of the transport coefficients
within the Kubo formalism. In section~\ref{GRT}, we derive the expressions of the
thermal widths to the $2\leftrightarrow 2$ and $1\leftrightarrow 2$ processes. Our numerical
results are presented in section~\ref{Res}.  Finally, section~\ref{Sum} present a summary
and present the main conclusions of the work. 

%
\section{Formalism}
\label{frmlsm}

We employ the Kubo formalism~\cite{{Kubo:1957mj},{Zubarev}} to compute the transport 
coefficients $\eta$ (shear viscosity), $\sigma$ (electical conductivity) and $\kappa$
(thermal conductivity). In this formalism, the transport coefficients are given in terms 
of correlation functions of operators involving components of the energy-momentum tensor 
$T_{\mu\nu}$, the quark-number $N_\mu$ and the electrical $J_\mu$ currents. Here, it is
assumed a system in the hydrodynamical regime, where relaxation time of the 
constituents is much shorter than the life time of the whole system. Under 
such an assumption, the medium is not far from equilibrium and in the evaluation of
statistical averages of the energy-momentum tensor and currents only linear terms in
spacetime gradients of local thermodynamical parameters (like temperature, velocity field, 
etc.) are retained; this is the linear response theory and leads to expressions for
transport coefficients that coincide with those derived within the relaxation 
time approximation~\cite{Chakraborty:2010fr}.

The energy-momentum tensor is given in terms of the Lagrangian density ${\cal L}$ of 
the model as
\be
T^{\mu\nu} = - g^{\mu\nu} {\cal L}
+ \frac{\del {\cal L}}{\del(\del_\mu\psi)}\del^\nu \psi
= g^{\mu\nu} {\cal L} + i{\ov \psi}\gamma^\mu\del^\nu\psi.
\ee
The quark-number $N_\mu$ and  electrical $J_\mu$ currents are given by
\be
N_\mu = \ov\psi\gamma_\mu\psi, \hspace{0.5cm}J_\mu = \ov\psi \hat Q \gamma_\mu \psi,
\ee
where $\hat Q$ is the charge matrix given in terms of the elementary electric charge
$e$ as
\be
\hat Q = e \left(
\begin{array}{cc}
2/3 & 0 \\
0   & -1/3 
\end{array}
\right).
\ee
The transport coefficients are given in terms of correlation functions of these quantities 
as
\be
\left(
\begin{array}{c}
\eta \\[0.2true cm]
\sigma \\[0.2true cm]
\kappa
\end{array}
\right) = \lim_{q_0 , |\bq| \rightarrow 0^+}  \; \frac{1}{q^0}
\left(
\begin{array}{c}
\frac{1}{20}A_\eta(q^0,\bq) \\[0.2true cm]
\frac{1}{6} A_\sigma(q^0,\bq) \\[0.2true cm]
\frac{-\beta}{6} A_\kappa(q^0,\bq)
\end{array}
\right),
\label{eta-zeta}
\ee
with $\beta = 1/T$, where $T$ is the temperature, and $A_\eta(q)$, $A_\sigma(q)$ and 
$A_\kappa(q)$ are the spectral functions: 
\bea
A_\eta(q) &=& \int d^4x \, e^{i q\cdot x} \,
\langle[\pi^{ij}(x),\pi^{ij}(0)]\rangle_\beta ,
\label{A_eta} \\
A_\sigma(q) &=& \int d^4x \, e^{i q \cdot x} \,
\langle[J_i(x),J^i(0)]\rangle_\beta ,
\label{A_sigma} \\
A_\kappa(q) &=& \int d^4x \, e^{i q \cdot x} \,
\langle[{\cal T}_i(x),{\cal T}^i(0)]\rangle_\beta ,
\label{A_kappa} \\
\eea
where
\bea
\pi^{ij}(x) &=& T^{ij}(x) - \frac{1}{3} \delta^{ij} T^{k}_{\,k}, 
\\[0.3true cm]
{\cal T}^i(x) &=& -T^{i 0}(x) - q \,N^i(x),
\eea
with $q$ being the enthalpy per particle, given in terms of the energy density $\epsilon$,
the pressure $P$ and net baryon density $\rho$ of the system as $q = (\epsilon + P)/\rho$. In Eqs.~(\ref{A_eta})-(\ref{A_kappa}), 
$\langle (..)\rangle_\beta$ denotes an appropriate thermal average. We are also interested 
in presenting results for the ratio $\eta/s$, where $s$ is the entropy density, given in terms
of $\epsilon$, $P$ and $\rho$ as $s= (\epsilon+P - \mu \rho)/T$, where $\mu$
is the baryon chemical potential.

We utilise the NJL model to derive the correlation functions and the thermodynamic functions. 
The Lagrangian density of the model for $u$ and $d$ flavors is given by~\cite{NJL-reviews}
\be
{\cal L} = {\ov \psi}(i\ds-m_Q)\psi
+ G \left[({\ov \psi}\psi)^2 + ({\ov \psi}i\gamma^5{\bm\tau}\psi)^2\right],
\label{LNJL}
\ee
where $m_Q$ is the current-quark mass matrix, which is diagonal with elements $m_u$ and $m_d$, 
and ${\bm \tau} = (\tau^1,\tau^2,\tau^3)$ are the flavor Pauli matrices. The model 
is solved in the quasi-particle approximation or, equivalently, in 
the leading-order approximation in the $1/N_c$ expansion, where $N_c = 3$ is the 
number of colors. This approximation is also equivalent to the traditional Hartree 
approximation of many-body theory. In this approximation, the thermal spectral 
functions and the thermodynamic functions are given in terms of the quark propagator. 

We employ the formalism of  real-time thermal field theory (RTF) to evaluate the 
correlation functions and thermodynamic functions. In RFT, the two point function 
of any field-theoretic operator has a $2 \times 2$ matrix structure reflecting the 
time ordering with respect to a contour in the complex plane~\cite{LeBellac}.
The relevant matrix can be diagonalized in terms of a single analytic function, 
which determines completely the dynamics of the corresponding two-point 
function {\textemdash} for details, see Ref.~\cite{G_Kubo}.  For example, when
neglecting dissipative effects, the $11$ component of the quark propagator in the 
quasi-particle approximation to the NJL model is given by 
\be
S_{11}(k) = (\slash\!\!\!k + M_Q) \, D_{11}(k),
\ee
with $k = (k_0, \bk)$ and
\begin{widetext}
\bea
D_{11}(k) &=& \frac{-1}{k^2_0 - (\omega^k_Q)^2 
+ i \varepsilon} - \, 2\pi i\omega^k_Q \delta(k^2_0 - (\omega^k_Q)^2) \, \left[n_Q(\bk) \theta(k_0) 
+ n_{\bar{Q}}(\bk) \theta(-k_0)\right],
\eea
where $\theta(k_0)$ is the step function, $\omega^k_Q = \left(\bk^2 + M^2_Q\right)^{1/2}$, 
with $M_Q$ being the constituent quark mass of a given flavor ($u$ or $d$), given by the solution of 
the gap equation~\cite{NJL-reviews}
\bea
M_Q &=& m_Q + 4 N_f N_cG\int^\Lambda_0 \frac{d^3\bk}{(2\pi)^3} \, \frac{M_Q}{\om^k_Q}
\, \left[1 - n_Q(\bk) - n_{\bar Q}(\bk)\right],
\label{gap}
\eea
where $N_f = 2$ and $N_c=3$ the numbers of flavor and colors, and $n_Q(\bk)$ and $n_{\bar Q}(\bk)$ are 
respectively the Fermi-Dirac distributions of quarks and antiquarks: 
\be
n_Q(\bk) = \frac{1}{e^{\beta(\om^k_Q - \mu)}+1},
\hspace{0.5cm}
n_{\bar Q}(\bk) = \frac{1}{e^{\beta(\om^k_Q + \mu)} + 1}.
\ee

Dissipative effects due to fluctuations introduce an imaginary part in the quark self energy 
giving a thermal width~$\Gamma_Q$, thereof modifying the quark propagator. Using the quark propagator 
modified by the width $\Gamma_Q$~\cite{{Hosoya},G_Kubo}, $\eta$, $\sigma$ 
and $\kappa$ can be readily obtained in the relaxation time approximation~\cite{Gavin,
Chakraborty:2010fr} or from the one-loop Kubo expression~\cite{Nicola_PRD,Nicola,G_Kubo}.
Their expressions are given by
\bea
\eta &=& \frac{2N_FN_c\beta}{15}
\int \frac{d^3\bk}{(2\pi)^3\Gamma_Q}\left(\frac{\bk^2}{\om^k_Q}\right)^2
\left\{{n_Q(\bk)}  \left[1-n_Q(\bk)\right]
+ n_{\bar Q}(\bk)\left[1-n_{\bar Q}(\bk)\right]
\right\} ,
\label{eta} 
\\
\sigma &=& \left(\frac{2N_c\beta}{3}\right) \left(\frac{5 e^2}{9}\right)
\int \frac{d^3\bk}{(2\pi)^3\Gamma_Q} \left(\frac{\bk}{\om^k_Q}\right)^2 
\left\{n_Q(\bk) 
\left[1 -n_Q(\bk)\right]
+ n_{\bar Q}(\bk) \left[1-n_{\bar Q}(\bk)\right]\right\} ,
\label{sigma}
\\
\kappa &=& \frac{2N_FN_c\beta^2}{3}\int \frac{d^3\bk}{(2\pi)^3\Gamma_Q}
\left(\frac{\bk}{\om^k_Q}\right)^2
\left\{ 
(\om^k_Q-q)^2 n_Q(\bk) \left[1-n_Q(\bk)\right] + (\om^k_Q+q)^2n_{\bar Q}(\bk)
\left[1-n_{\bar Q}(\bk)\right]
\right\} .
\label{kappa}
\eea
The explicit expression for the enthalpy $h$ in the present model, together with those
for other thermodynamic functions, are given the appendix.  The calculation of the contribution of 
quark-meson fluctuations to $\Gamma_Q$ requires meson masses $m_M$ and and quark-meson couplings 
$g_{MQQ}$, for $M=\sigma, \pi$. Their expressions in the quasi-particle approximation are well known 
in the literature~\cite{NJL-reviews}, but repeat them here for completeness and setting the notation: 
\bea
1 - 2 G \Pi_M(\omega^2=m^2_M) = 0, \hspace{1.0cm}
g^2_{MQQ} = \left[\frac{\partial \Pi_M(\omega^2)}{\partial \omega^2} 
\right]^{-1}_{\omega^2 = m^2_M} ,
\label{mM-gQQM}
\eea 
where $\Pi_M(\omega^2)$ is the proper polarization function
\bea
\Pi_M(\omega^2) = 2N_c N_f \int^\Lambda_0 \frac{d^3\bk}{(2\pi)^3} 
\frac{F_M(\omega^2)}{\omega^k_Q} \, \left[1-n_Q(\bk)-n_{\bar Q}(\bk)\right],
\label{PiM}
\eea
with
\bea
F_\pi(\omega^2) = \frac{(\omega^k_Q)^2}{(\omega^k_Q)^2 - \omega^2/4}, 
\hspace{1.0cm}
F_\sigma(\omega^2) = \frac{(\omega^k_Q)^2 - M^2_Q}{(\omega^k_Q)^2 - \omega^2/4} .
\label{F-functions}
\eea
The integrals in Eq.~(\ref{PiM}) are evaluated as  principal-value integrals 
when $\omega^2 > 4 M^2_Q$. 

\end{widetext}

For the same reasons, we present the expressions for the pressure~$P$, the energy 
density~$\epsilon$, and baryon density~$\rho$:
\bea
P &=& 2N_fN_c\int \frac{d^3k}{(2\pi)^3}\frac{\bk^2}{3\om^k_Q}
\left[n_Q(\om^k_Q)+n_{\bar Q}(\om^k_Q)\right],
\label{P} \\
\ep &=& 2N_fN_c\int \frac{d^3k}{(2\pi)^3} \om^k_Q
\left[n_Q(\om^k_Q)+n_{\bar Q}(\om^k_Q)\right],
\label{epsilon} \\
\rho &=& 2N_fN_c\int \frac{d^3k}{(2\pi)^3}\left[n_Q(\om^k_Q)-n_{\bar Q}(\om^k_Q)\right].
\label{rho_Tmu}
\eea
The entropy density~$s$ and enthalpy density~$h$ are related to the above quantities through
the following relations:
\bea
s &=& \frac{\ep + P-\mu \rho}{T},
\label{s_eP} \\[0.3true cm]
h &=& \ep + P .
\label{h_eP}
\eea
Another important thermodynamical quantity is the heat function $q$ for each 
quark, defined by the ratio of enthalpy density to the net quark density, 
$q = h/\rho$. This quantity becomes divergent (unphysical) at $\mu=0$, where 
net quark density vanishes.

%
\section{Computation of thermal widths}
\label{GRT}

We evaluate the thermal width $\Gamma_Q$ including contributions from $2 \leftrightarrow 2$ 
and $1\leftrightarrow 2$ types of processes. The $2 \leftrightarrow 2$ processes refer to 
in-medium quark-antiquark and antiquark-antiquark scatterings mediated by $\sigma$ and $\pi$ 
exchanges, denoted generically by $QQ \leftrightarrow QQ$. The $1\leftrightarrow 2$ processes
refer to quark-meson fluctuations, denoted by $Q \leftrightarrow QM$.  

For the two-flavor case, there are twelve possible $QQ \leftrightarrow QQ$ processes, whose matrix 
elements $\ov {M}_{QQ \leftrightarrow QQ}$ are written down explicitly in 
Refs.~\cite{Deb:2016myz,klevansky,klevansky2,{Abhishek:2017pkp}}, 
which we use for calculating the $\Gamma_{QQ\leftrightarrow QQ}$ contribution to to the full width.
Let us assume $k$ and $p$ ($k'$ and $p'$) for the initial (final) four-momenta in the scattering 
processes $Q(k) + Q(p)\rightarrow Q(k') + Q(p')$. Hence, the collisional width of a probe particle 
with momentum $\bk$ will be a function of the temperature $T$ and chemical potential $\mu$ of the 
medium, with the momenta $\bp$, $\bk'$, $\bp'$ of the remaining participating
particles in the scattering are integrated out, can be written as
\begin{widetext}
\bea
\Gamma_{QQ\leftrightarrow QQ}(\bk,T,\mu)&=& \int \frac{d^3\bp}{(2\pi)^3 2\om^p_Q } 
\frac{n_{Q/\bar Q}(\bp)}{1+\delta_{kp}} 
\int \frac{d^3\bk'}{(2\pi)^3 2\om^{k'}_Q } 
\int \frac{d^3\bp'}{(2\pi)^3 2\om^{p'}_Q } \, 
\left[1-n_{Q/\bar Q}(\bk')\right] \, \left[1-n_{Q/\bar Q}(\bp')\right] \nn \\[0.3true cm]
&& \times \,  (2\pi)^4\delta^4(k+p-k'-p') \, |\ov {M}|^2_{QQ \leftrightarrow QQ} ,
\label{gm_Sc}
\eea
%
where $1 + \delta_{k,p}=2(1)$ for identical (nonidentical) quarks/antiquarks in the initial 
and final states. As in Ref.~\cite{Abhishek:2017pkp}, we include a finite thermal 
width in the meson propagators in the expressions for $|\ov {M}|^2$.

Next, we consider the inelastic processes $Q \leftrightarrow QM$, in which a quark/antiquark 
can emit or absorb a thermalized thermalized meson from the medium. Their contributions 
$\Gamma_{Q\leftrightarrow QM}$ can be obtained from the Landau cut part of quark 
self-energy coming from quark-meson loops~\cite{Weldon}. To evaluate the quark self-energy, 
we employ the quark-meson interaction Lagrangian densities~\cite{Quack_Klevansky},
\bea 
{\cal L}_{\pi QQ} &=& ig_{\pi QQ}\sum_{f=u,d} \bar{\psi}_f \,\gamma^5 \, 
\btau \cdot \bpi \, \psi_f ,
\label{LQQpi} \nn \\
{\cal L}_{\sigma QQ} &=& g_{\sigma QQ} \sum_{f=u,d} \bar{\psi}_f \, \sigma \, \psi_f ,
\label{LQQsig}
\eea
where the quark-meson couplings are obtained in the NJL model from Eq.~(\ref{mM-gQQM}). Given
these, the imaginary part of the quark self-energy can be evaluated. Analyzing the branch cuts 
of this quark self-energy at finite temperature, one can easily find the quark pole
$k = (\om^k_Q,\bk)$ within the Landau-cut region $\bk <k_0< [\bk^2+(M_Q-m_M)^2]^{1/2}$ 
for $m_M>2M_Q$, and write for $\Gamma_{Q\leftrightarrow QM}$~\cite{G_Kubo}:
%
\bea
\Gamma_{{Q\leftrightarrow QM}} (\bk,T,\mu) &=& - \frac{1}{2M_Q}
{\rm Tr}\left[(\ks +M_Q){\rm Im}{\Sigma}^R_{Q(QM)}(k)\right]_{k_0=\om^k_Q}
\nn\\
&=& \left[\int\frac{d^3{\bl}}{(2\pi)^3} \, \delta(k^0 + \omega^l_Q - \om^u_M) \, 
\frac{n_Q(\bl) + n_M(\bu)}{4 \om^l_Q \om^u_M} \,
L_{Q(QM)}(l^0 = - \omega^l_Q,\bl,k)\right]_{k^0=\om^k_Q}~,
\label{gm_Im}
\eea
\end{widetext}
where $\bu = \bk - \bl$, $n_M(\bu)$ is the Bose-Einstein distribution for mesons 
with energy $\om^u_M = (\bu^2 + m^2_M)^{1/2}$:
\bea
n_M(\bk) = \frac{1}{e^{\beta\om^u_{M}}-1} ,
\eea
and 
\bea
L_{Q(Q\pi)}(l,k) &=& 3\frac{4g_{\pi}^2}{2M_Q}\left[M_Q^2-(k\cdot l)\right],
\nn \\
L_{Q(Q\sigma)}(l,k)&=& \frac{4g_{\sigma}^2}{2M_Q}\left[M_Q^2+(k\cdot l)\right] .
\eea
The contributions from antiquarks are obtained from Eq.~(\ref{gm_Im}) by
replacing $n_Q(\om^l_Q)$ by $n_{\bar Q}(\om^l_Q)$. Finally, we define 
average thermal widths $\Gamma_{QQ\leftrightarrow QQ}$ and $\Gamma_{Q(QM)}$ by
the averages of $\Gamma_{QQ\leftrightarrow QQ}(\bk,T,\mu)$ and 
$\Gamma_{{Q\leftrightarrow QM}} (\bk,T,\mu)$ over the thermal distributions:
\be
\Gamma(T,\mu) = \frac{
\int \frac{d^3\bk}{(2\pi)^3} \Gamma(\bk,T,\mu) \, n_{Q/{\bar Q}}(\bk)
}
{\int \frac{d^3\bk}{(2\pi)^3} \, n_{Q/{\bar Q}}(\bk)}~.
\label{th_av}
\ee
%

%
\section{Results}
\label{Res}

\begin{figure} 
\includegraphics[scale=0.65]{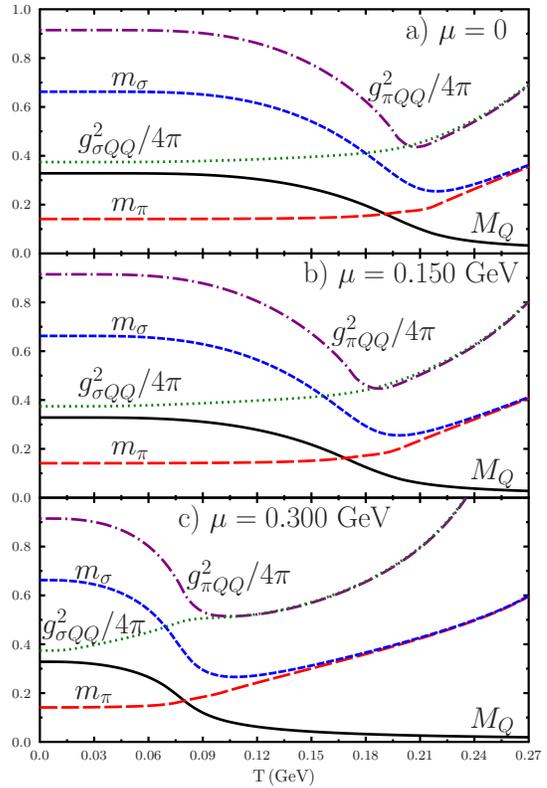}
\caption{The constituent quark mass $M_Q$ (black solid line), pion mass 
$m_\pi$ (red long dashed line), sigma mass $m_\sigma$ (blue short dashed line)
in {\rm GeV}, and the quark-meson couplings $g^2_{\sigma QQ}/4\pi$ (green dotted line) 
and $g^2_{\pi QQ}/4\pi$ (purple dash-dotted line) as a function of $T$ for 
three different values of chemical potential $\mu$. }
\label{MWC_T}
\end{figure}

\begin{figure} 
\includegraphics[scale=0.35]{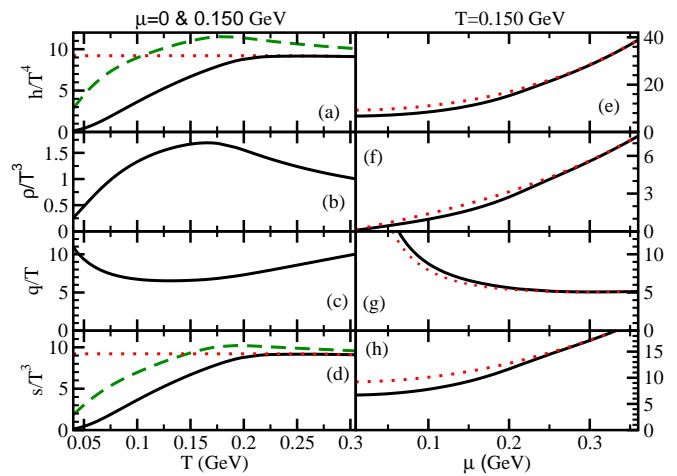}
\caption{Dimensionless ratios to the appropriate powers of $T$ of the 
thermodynamical functions $h$, $\rho$, $q$, and $s$. Left panel: $T$ dependence for
$\mu = 0$ (dashed green) and $\mu = 0.15~{\rm GeV}$ (solid black). 
Right Panel: $\mu$ dependence for $T = 0.15~{\rm GeV}$. Dotted red curves are 
the ratios calculated with $M_Q = 0$ (on the left panel, $\mu = 0$).}
\label{hs_Tmu}
\end{figure}

The parameters of the model are fixed to obtain realistic values for the quark 
condensate $\langle \bar uu\rangle = \langle \bar dd\rangle = (-252~\rm{MeV})^3$, 
pion leptonic decay constant $f_\pi = 94$~MeV and the pion mass $m_\pi = 142$~MeV. 
The parameters to be fixed are the current quark masses $m_Q = (m_u,m_d)$, the 
coupling $G$ and the cutoff mass $\Lambda$. In present calculation, they are fixed 
to $m_Q = m_u = m_d = 5$~MeV, $G\Lambda^2 = 2.14$, and $\Lambda = 653$~MeV. 
At $T=0$ and $\mu=0$, the constituent quark and $\sigma$ meson masses are 
$M_Q = M_u = M_d = 328$~MeV and $m_\sigma = 663$~MeV. For completeness and clarity 
of presentation of our results on the transport coefficients, we present in Fig.~\ref{MWC_T} 
the $T$ and $\mu$ dependences of the quark and meson masses and quark-meson couplings,
and in Fig.~\ref{hs_Tmu} we present the thermodynamical functions .

Figure~\ref{MWC_T} reveals the well known facts that while the constituent quark mass
$M_Q$ and the $m_\sigma \simeq 2 M_q$ and $g_{\pi QQ}$ drop significantly up to a 
pseudocritical temperature, the $\pi$ mass and $g_{\sigma QQ}$ change very little with
$T$. In addition, the pseudocritical temperature decreases when $\mu$ increases, reflecting 
the fact that both $T$ and $\mu$ lead to a partial restoration of chiral symmetry. For values 
of $T$ sufficiently higher than the pseudocritical temperature, the meson masses and quark-meson
couplings become degenerate, reflecting the restoration of the approximate chiral symmetry 
of the Lagrangian.

Figure~\ref{hs_Tmu} presents the results for the thermodynamical functions 
$h$, $\rho$, $q$ and $s$, normalized by the appropriate powers of $T$ to obtain
dimensionless ratios. To emphasize the effect
of dynamical chiral symmetry breaking on those functions, they are 
also shown (red dotted lines) for massless quarks, i.e. they 
are calculated by setting $M_Q = 0$ in Eqs.~(\ref{P})-(\ref{h_eP}).
Clearly, at low $(T,\mu)$ values the effect of symmetry breaking is 
substantial and at high $T$ one has the recovery of the Lagrangian symmetry.
The heat function $q=h/\rho$ increases substantially for small values of $\mu$ 
because the net quark density $\rho$ decreases as $\mu \rightarrow 0$. The thermal 
conductivity $\kappa$, being proportional $q$, increases substantially for small 
$\mu$ as well.

\begin{figure} 
\includegraphics[scale=0.35]{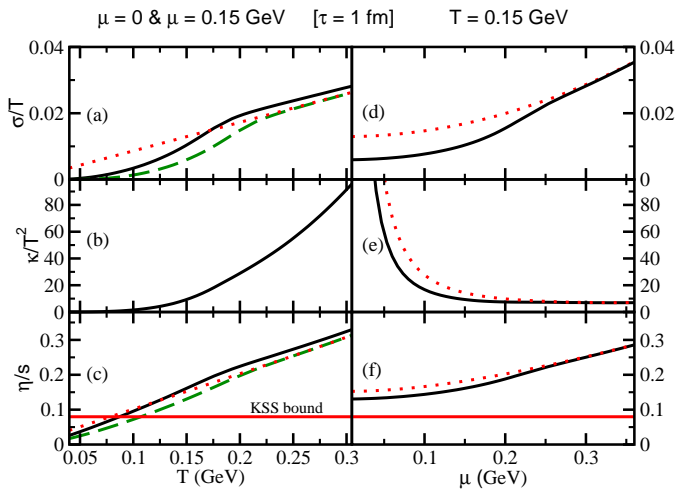}
\caption{$T$ dependence of normalized electrical conductivity $\sigma/T$ (a), 
thermal conductivity $\kappa/T^2$ (b) and shear viscosity $\eta/s$ (c). Their $\mu$ dependence
at $T=0.150$ GeV are respectively in (d), (e) and (f). All results are obtained for 
$\tau = 1/\Gamma =1$~fm. The (red) dotted lines are the results for massless quarks and
the horizontal (red) solid line indicates the KSS holographic bound, $\eta/s= 1/4\pi$.}
\label{tr_Tmu}
\end{figure}

\begin{figure} 
\includegraphics[scale=0.35]{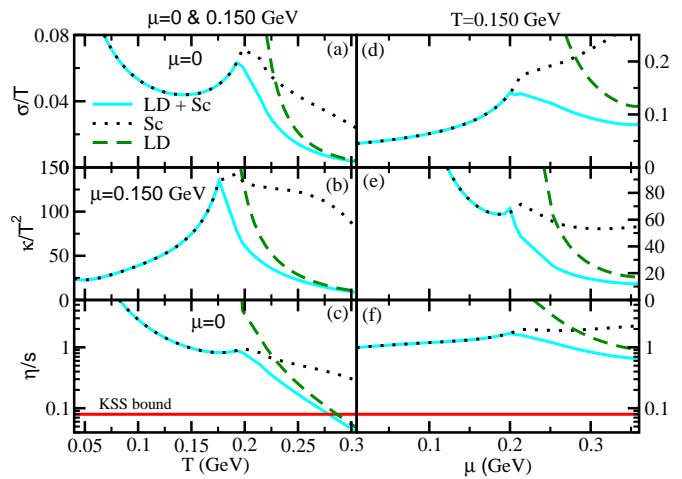}
\caption{(Color on-line) $T$ dependence of $\sigma/T$ (a), 
$\kappa/T^2$ (b) and $\eta/s$ (c). Their $\mu$ dependence
at $T=0.150$ GeV are respectively in (d), (e) and (f). Contributions
of LD (dashed line), Sc (dotted line) components and their total (solid line)
in these transport coefficients are individually shown.}
\label{trGQ_Tmu}
\end{figure}

Next, we present our results for the transport coefficients. To get insight into the
importance of using a thermal width $\Gamma_Q(\bk,T,\mu)$ in which the $QQ \leftrightarrow QQ$ 
and $Q \leftrightarrow QM$ physical processes are treated consistently with the
dynamics of chiral restoration, let us initially contrast results for the transport 
coefficients when one uses a $(T,\mu)$-independent thermal width in Eqs.~(\ref{eta})-(\ref{kappa}). 
We choose $\Gamma \equiv 1~{\rm fm}^{-1}${\textemdash}we often refer to the relaxation time, which 
is the inverse of the thermal width, $\tau = 1/\Gamma$. The results for a $(T,\mu)-$independent 
$\Gamma(\bk,T,\mu)$ are shown in Fig.~\ref{tr_Tmu}, while those with the full $(T,\mu)$ dependence 
of $\Gamma(\bk,T,\mu)$ are shown in Fig.~\ref{trGQ_Tmu}. Clearly, they are markedly different. 
While the results in Fig.~\ref{tr_Tmu} are determined solely by phase space, those in Fig.~\ref{trGQ_Tmu} 
feature the interplay between smooth contributions from the $QQ \leftrightarrow QQ$ scattering processes 
(Sc, dotted lines) and Landau-cut cusp contribution from the $Q \leftrightarrow QM$ processes 
(LD, dashed lines). The results are easily understood examining in detail the Sc and LD contributions
to the thermal widths, which we discuss next.

\begin{figure} 
\includegraphics[scale=0.35]{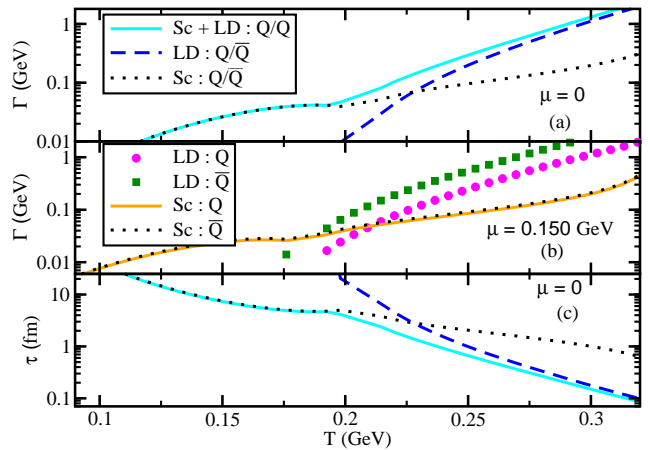}
\caption{(Color on-line) Contribution of Landau damping (LD) part (dashed line), 
$2\leftrightarrow 2$ scattering part (solid line)
and their total (dotted line) in the thermal widths (a) and relaxation times (c) of 
quark/anti-quark (Q/${\ov Q}$) at $\mu=0$. (b) at $\mu=0.150$ GeV, circles and squares 
stand for LD part of thermal width for Q and ${\ov Q}$, while their scattering parts 
are plotted by solid and dotted lines.}
\label{GQ_T}
\end{figure}

\begin{figure} 
\includegraphics[scale=0.35]{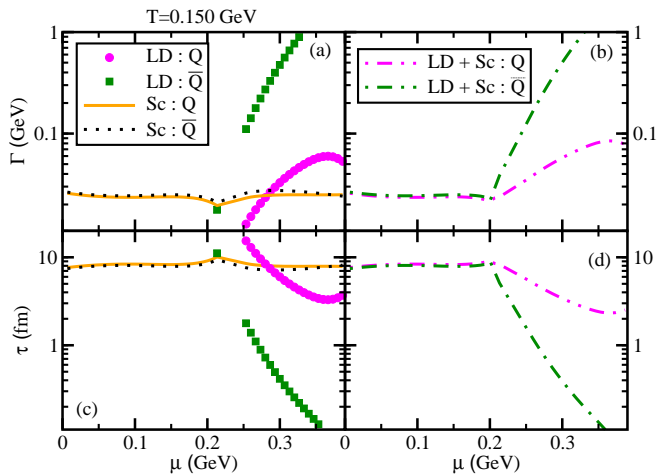}
\caption{(Color on-line) $\mu$ dependence of thermal widths (a) and relaxation times (c)
for LD and scattering parts of quark and anti-quark. Their total thermal widths (b) 
and relaxation times (d) are plotted by dash-double dotted and dash-dotted lines.}
\label{GQ_mu}
\end{figure}

The temperature dependence of $\Gamma$ and of its inverse $\tau$ is shown in Fig.~\ref{GQ_T} 
for two values of the chemical potential, $\mu = 0$ and $\mu = 0.15$~GeV. As can be seen in
the figure, while the scattering (Sc) contribution is a smooth function of the temperature, 
the Landau-cut (LD) features a cusp at a critical temperature $T_M$, the Mott temperature.
This is a threshold temperature beyond which $m_\pi(T\geq T_M)\geq 2M_Q(T\geq T_M)$, when 
the pionic bound state delocalizes into its constituents.  The Mott temperature, although closely 
related to the chiral pseudocritical temperature discussed before, in the present 
model it is larger than the latter. In the chiral restored phase the system consists of 
a mixture of quarks and antiquarks and pions and for $T > T_M$ the pions as bound states 
disappear{\textemdash}for a thorough discussion on these temperatures, and
their relation to the one of quark deconfinement, see e.g. Refs.~\cite{klevansky2,{Hansen:2006ee},
{Costa:2008dp},{Blaschke:2014zsa}}. Therefore, below $T_M$, the LD contribution to the thermal 
width is zero. Beyond $T_M$, the thermal width gets strongly enhanced, meaning that 
that quarks and antiquarks quickly thermalize in the medium. The figure also reveals that 
$T_M$ decreases with $\mu$. One can also 
identify a Mott chemical potential $\mu_M$, as shown in Fig.~\ref{GQ_mu}. For 
$T = 0.150$~GeV, $\mu_M \sim 0.2$~GeV. Notice also that while the quark and anti-quark Sc 
and LD contribuions are equal for $\mu = 0$, they are different at finite $\mu$, a feature 
that is obviously due to the different dependence with $\mu$ of the quark and antiquark 
Fermi-Dirac distributions. We note that both quark and antiquark have the same $T_M$ and 
$\mu_M$, but they seem different in Figs.~\ref{GQ_T}(b) and Fig.~\ref{GQ_mu}(a) because the 
quark contribution becomes too small close to threshold to become visible in those graphs.
It is important to note that a cusp structure might not be present when using different
set of model parameters which predict different values for the constituent-quark 
masses and quark-meson couplings. For example, Refs.~\cite{klevansky,Redlich_NPA,Deb:2016myz} 
found a valley structure near the Mott transition temperature instead of a cusp structure.

Let us return to the transport coefficients. The $\eta/s$ curve in Fig.~\ref{trGQ_Tmu} 
reveals interesting features. Below the chiral pseudocritical temperature, the magnitude 
of $\eta/s$ is much larger than the KSS bound~\cite{Kovtun:2004de}, $\eta/s = 1/4\pi$,
shown by the horizontal red line in that figure, indicating that matter at those temperatures 
may not at all behave like a perfect fluid. However, beyond Mott temperature, $\eta/s$ is abruptly 
reduced and approaches to KSS bound, crossing that bound at $T\sim 0.275$~GeV. The main cause 
for this behavior is the LD the contribution, which dominates the Sc contribution 
beyond the Mott temperature. We tress that at those high temperatures, there migh
exist additional contributions coming from gluonic degrees of freedom, which are
not explicitly taken into account by the model and might raise the value of
$\eta/s$. In the context of the present model, the message is that the LD 
contribution due to $Q\leftrightarrow QM$ inelasticscatterings in the medium 
is the origin of perfect fluid nature of quark matter near and above Mott temperature. 
The same is true for the electrical and thermal conductivities, they are also
small for high values of $T$ and $\mu$ due a lower relaxation time of quarks due 
to inelastic scatterings.

%
\section{Summary and Conclusions} 
\label{Sum}

In this work we have made a comparative study of the relative contributions of 
elastic $2 \leftrightarrow 2$ and inelastic $1 \leftrightarrow 2$ scatterings 
in electrical and thermal conductivities and shear viscosity of strongly interacting 
matter in the context of the NJL model. This is the first study taking into account both 
types of scatterings; previously, 
Refs.~\cite{Ghosh:2015mda,G_CAPSS,Weise2,LKW,Harutyunyan:2017ttz,HMPQM1,klevansky,klevansky2,Marty,Deb:2016myz}
of effective QCD model calculations, Refs.~\cite{Ghosh:2015mda,G_CAPSS,Weise2,LKW,Harutyunyan:2017ttz,HMPQM1} 
have estimated $1\leftrightarrow 2$ type scattering like 
quark $\leftrightarrow$ quark + meson by calculating quark self-energy for quark-meson
loop at finite temperature. These $1\leftrightarrow 2$ type scattering contribute
within a very narrow temperature window, which is near but above Mott 
temperature. On the other hand, Refs.~\cite{klevansky,klevansky2,Marty,Deb:2016myz} have 
investigated the contributions of $2\leftrightarrow 2$ type scattering of quark in the transport
coefficients, which contribute in entire temperature range. A simultaneous role 
of $2\leftrightarrow 2$ and $1\leftrightarrow 2$ type scatterings
on transport coefficients has never been studied and the present work has
provides such an investigation.

In the language of the Kubo formalism, the $T$ and $\mu$ dependence of transport coefficients 
has two sources in this model. One is thermodynamical phase-space via the Fermi-Dirac
distribution functions, which have explicit and implicit through the quark masses
$T$ and $\mu$ dependences. The other is via the thermal width, which is calculated 
here self-consistently with the dynamics of chiral restoration taking into account 
elastic and inelastic scatterings. The self-consistency has a dramatic effect on the
transport coefficients, as we demonstrated by comparing the self-consistent results 
with those obtained with a constant thermal width.
In this case, shear viscosity to entropy density ratio $\eta/s$
and electrical conductivity to temperature ratio $\sigma/T$ increase with both 
$T$ and $\mu$. Their rate of increments are changed when one approaches
from hadron to quark phases in $T$-$\mu$ plane.
Owing to the definition, thermal conductivity is generally diverged at $\mu=0$
but its divergence is removed for finite $\mu$. It rapidly decreases with $\mu$
because of thermodynamical quantity, enthalpy to net quark density ratio $h/\rho$,
and then after $\mu=0.150$ GeV, it remain more or less constant, whose strength
is proportionally determined by relaxation time.

Next, an explicit $T$ and $\mu$ dependence of thermal
width of quark has been calculated from different quark-quark, quark-anti-quark 
elastic scatterings via meson exchanges and quark-meson in-elastic scatterings.
All are in-medium scatterings and similarly, one can calculate
anti-quark relaxation time by considering suitable diagrams. 
The in-elastic scatterings are estimated 
from imaginary part of quark self-energy due to quark-meson loops.
Due to Mott effect, the quark meson in-elastic scattering has certain $T$-$\mu$ threshold,
beyond which it becomes non-zero. However, elastic scatterings provide
non-zero relaxation time in entire $T$-$\mu$ plane. Along the Mott
curve or $T_M$-$\mu_M$ curve, it carry a mild cusp structure,
which is also reflected in the $(T, \mu)$ profile of transport coefficients.
Adding elastic and in-elastic scatterings, we get total relaxation time
of quark and anti-quark, for which we get very small $\eta/s$, close
to its KSS bound. However, this possibility is expected near and above Mott curve
in $T$-$\mu$ plane, where in-elastic scatterings suddenly blow up. 
Within this $T$-$\mu$ window, our outcome
is supporting the picture of perfect fluid nature, observed in RHIC matter.
Due to this lower relaxation time in this  $T$-$\mu$ window,
the other transport coefficients like electrical and thermal conductivities
will also be small.

{\bf Acknowledgment:} Work partially financed by by Conselho Nacional de Desenvolvimento 
Cient\'{\i}fico e Tecnol\'{o}gico - CNPq, 305894/2009-9 (G.K.), 464898/2014-5(G.K) (INCT F\'{\i}sica 
Nuclear e Apli\-ca\-\c{c}\~oes), 168240/2017-3 (F.E.S), and Funda\c{c}\~{a}o de Amparo \`{a} 
Pesquisa do Estado de S\~{a}o Paulo - FAPESP, 2013/01907-0 (G.K).
SG, AA, HM acknowledge Workshop in High Energy Physics Phenomenology (WHEPP), 2017 for getting
some fruitful discussions on this work.


\begin{thebibliography}{99}
%
\bibitem{Buballa:2014jta} 
  M.~Buballa, V. Dexheimer, A. Drago, E. Fraga, P. Haensel, I. Mishustin, 
  G. Pagliara, J. Schaffner-Bielich, S. Schramm, A. Sedrakian and F. Weber,
  J.\ Phys.\ G {\bf 41}, no. 12, 123001 (2014).
%
\bibitem{Baym:2017whm} 
  G.~Baym, T.~Hatsuda, T.~Kojo, P.~D.~Powell, Y.~Song and T.~Takatsuka,
  Rept.\ Prog.\ Phys.\  {\bf 81}, no. 5, 056902 (2018).
%
\bibitem{Bratkovskaya:2017ssq} 
  E.~L.~Bratkovskaya, W.~Cassing, P.~Moreau and T.~Song,
  KnE Energ.\ Phys.\  {\bf 3}, 234 (2018).
%
\bibitem{Shuryak:2014zxa} 
  E.~Shuryak,
  Rev.\ Mod.\ Phys.\  {\bf 89}, 035001 (2017).
%
\bibitem{Rapp_rev} R. Rapp
Adv. High Energy Phys. 2013, 148253 (2013);
R. Rapp, J. Wambach , Adv. Nucl. Phys. 25, 1 (2000).
%
\bibitem{G_rev} P. Mohanty, S. Ghosh, S. Mitra
Adv. High Energy Phys. 2013, 176578 (2013).
%
\bibitem{SPS} R. Arnaldi et al. Phys. Rev. Lett. {\bf 100}, 
022302 (2008); R. Arnaldi et al., Eur. Phys. J. C {\bf 61}, 
711 (2009); S. Damjanovic et al., J. Phys. G {\bf 35}, 
104036 (2008).
%
%
\bibitem{LQCD_Buividovich} P. V. Buividovich, M. N. Chernodub, D. E. Kharzeev,
T. Kalaydzhyan, E. V. Luschevskaya, and M. I. Polikarpov,
Phys. Rev. Lett. 105, 132001 (2010).
%
\bibitem{LQCD_Amato}A. Amato, G. Aarts, C. Allton, P. Giudice, S. Hands, J.I. Skullerud, 
Phys. Rev. Lett. 111, 172001 (2013).
%
\bibitem{LQCD_Burnier}Y. Burnier and M. Laine, 
Eur. Phys. J. C 72, 1902 (2012).
%
\bibitem{LQCD_Arts_2007}G. Aarts, C. Allton, J. Foley, S. Hands, and S. Kim, 
Phys. Rev. Lett. 99, 022002 (2007).
%
\bibitem{LQCD_Barndt}B. B. Brandt, A. Francis, H. B. Meyer, and H. Wittig,
J. High Energy Phys. 03 (2013) 100.
%
\bibitem{LQCD_Ding}H.T. Ding, A. Francis, O. Kaczmarek, F. Karsch, 
E. Laermann, and W. Soeldner, 
Phys. Rev. D 83, 034504 (2011).
%
\bibitem{LQCD_Gupta}S. Gupta, 
Phys. Lett. B 597, 57 (2004).

\bibitem{Ratti:2018ksb} 
  C.~Ratti,
  Rept.\ Prog.\ Phys.\  {\bf 81}, no. 8, 084301 (2018).
%
\bibitem{Nicola} D. Fernandez-Fraile and A. Gomez Nicola,
Eur. Phys. J. C {\bf 62}, 37 (2009).
%
\bibitem{Marty} R. Marty, E. Bratkovskaya, W. Cassing, J. Aichelin, 
and H. Berrehrah, Phys. Rev. C {\bf 88}, 045204  (2013) .
%
%
\bibitem{Gavin} S. Gavin, 
Nucl. Phys. A {\bf 435}, 826 (1985).
%
\bibitem{Prakash} M. Prakash, M. Prakash, R. Venugopalan, and G. Welke, 
Phys. Rep. {\bf 227}, 321 (1993).

%
\bibitem{Deb:2016myz} 
  P.~Deb, G.~P.~Kadam, and H.~Mishra,
  Phys.\ Rev.\ D {\bf 94}, 094002 (2016).
%
\bibitem{Ghosh:2015mda} 
  S.~Ghosh, T.~C.~Peixoto, V.~Roy, F.~E.~Serna, and G.~Krein,
  Phys.\ Rev.\ C {\bf 93}, 045205 (2016).
%
\bibitem{G_CAPSS}
S. Ghosh, A. Lahiri, S. Majumder, R. Ray, S. K. Ghosh, 
Phys. Rev. C {\bf 88}, 068201 (2013).
%
\bibitem{Weise2}
R. Lang and W. Weise
Eur. Phys. J. A {\bf 50}, 63 (2014).
%
\bibitem{LKW} 
R. Lang, N. Kaiser, and W. Weise,
Eur. Phys. J. A {\bf 51}, 127 (2015).
%
\bibitem{klevansky} 
P. Zhuang, J. Hufner, S.P. Klevansky, and L. Neise,
Phys. Rev. D {\bf 51}, 3728 (1995).
%
\bibitem{klevansky2} 
P. Rehberg, S.P. Klevansky, and J. Hufner,
Nucl. Phys. A {\bf 608}, 356 (1996).
%
\bibitem{Redlich_NPA} C. Sasaki and K. Redlich,
Nucl. Phys. A {\bf  832}, 62 (2010).
%
%
\bibitem{Abhishek:2017pkp} 
  A.~Abhishek, H.~Mishra, and S.~Ghosh,
  Phys.\ Rev.\ D {\bf 97}, 014005 (2018).
%
%
\bibitem{Harutyunyan:2017ttz} 
  A.~Harutyunyan, D.~H.~Rischke and A.~Sedrakian,
  Phys.\ Rev.\ D {\bf 95}, no. 11, 114021 (2017).
  
\bibitem{HMPQM1}P. Singha, A. Abhishek, G. Kadam, S. Ghosh, and H. Mishra
arXiv:1705.03084 [nucl-th].
%
  
%
\bibitem{Lee} C. Lee and I. Zahed,
Phys. Rev. C {\bf 90}, 025204 (2014).
%
\bibitem{Nicola_PRD} D. Fernandez-Fraile and A. Gomez Nicola,
Phys. Rev. D {\bf 73}, 045025 (2006).
%
\bibitem{G_El} S. Ghosh,
Phys. Rev. {\bf D 95} (2017) 036018
%
\bibitem{Cassing} W. Cassing, O. Linnyk, T. Steinert, and V. Ozvenchuk,
Phys. Rev. Lett. {\bf 110}, 182301 (2013).
%
\bibitem{Finazzo} S. I. Finazzo and J. Noronha
Phys. Rev. D {\bf 89}, 106008 (2014).
%
\bibitem{Puglisi} A. Puglisi, S. Plumari, and V. Greco,
Phys. Rev. D {\bf 90}, 114009 (2014);
J. Phys. Conf. Ser. {\bf 612}, 012057 (2015);
Phys. Lett. B{\bf 751}, 326 (2015).
%
\bibitem{Greif} M. Greif, I. Bouras, Z. Xu, and C. Greiner,
Phys. Rev. D {\bf 90}, 094014  (2014);
J. Phys. Conf. Ser. {\bf 612}, 012056 (2015) .
%
%
%
\bibitem{Davesne} D. Davesne, 
Phys. Rev. C {\b 53}, 3069 (1996).
%
\bibitem{Nam} S. Nam,
Mod. Phys. Lett. A {\bf 30}, 1550054 (2015).
%
%
\bibitem{NJL_Iwasaki} M. Iwasaki and T. Fukutome, 
J. Phys. G {\bf 36}, 115012 (2009).
%
%
%
%
\bibitem{Sarkar} S. Mitra and S. Sarkar, 
Phys. Rev. D 89 (2014) 054013;
S. Mitra, U. Gangopadhyaya and S. Sarkar, 
Phys. Rev. D 91 (2015) 094012.
%
%
\bibitem{CFL1} I. A. Shovkovy and P. J. Ellis, 
Phys. Rev. C {\bf 66}, 015802 (2002).
%
\bibitem{CFL2} M. Braby, J. Chao, and T. Sch\"afer, 
Phys. Rev. C {\bf 81}, 045205 (2010).
%
%
%
\bibitem{Kubo:1957mj} 
  R.~Kubo,
  J.\ Phys.\ Soc.\ Jap.\  {\bf 12}, 570 (1957).
%
\bibitem{Zubarev} D. N. Zubarev
{\it Non-equilibrium statistical thermodynamics}
(New York, Consultants Bureau, 1974).
%
\bibitem{Chakraborty:2010fr} 
  P.~Chakraborty and J.~I.~Kapusta,
  Phys.\ Rev.\ C {\bf 83}, 014906 (2011).
%
%
\bibitem{LeBellac} 
M. Le Bellac, {\it Thermal Field Theory} (Cambridge University Press, 
Cambridge, England, 2000).
%
\bibitem{G_Kubo} S. Ghosh,
Int. J. Mod. Phys. A {\bf A 29}, 1450054 (2014).
%
%
\bibitem{NJL-reviews} 
U. Vogl and W. Weise, Prog. Part. Nucl. Phys. {\bf 27}, 195 (1991);
S. P. Klevansky, Rev. Mod. Phys. {\bf 64}, 649 (1992);
T. Hatsuda and T. Kunihiro, Phys. Rep. {\bf 247}, 221 (1994);
M. Buballa, Phys. Rep. {\bf 407}, 205 (2005).
%
\bibitem{Hosoya} 
A. Hosoya, M.-A. Sakagami, and M. Takao,
Ann. Phys. (N.Y.) {\bf 154}, 229 (1984).
%
\bibitem{Kovtun:2004de} 
  P.~Kovtun, D.~T.~Son and A.~O.~Starinets,
  Phys.\ Rev.\ Lett.\  {\bf 94}, 111601 (2005).
%
\bibitem{Weldon} H.A. Weldon, 
Phys. Rev. D {\bf D 28}, 2007 (1983).
%
\bibitem{Quack_Klevansky} E. Quack, S. P. Klevansky,
Phys. Rev. {\bf C 49}, 6 (1994). 
%
\bibitem{Hansen:2006ee} 
  H.~Hansen, W.~M.~Alberico, A.~Beraudo, A.~Molinari, M.~Nardi and C.~Ratti,
  Phys.\ Rev.\ D {\bf 75}, 065004 (2007).

%
\bibitem{Costa:2008dp} 
  P.~Costa, M.~C.~Ruivo, C.~A.~de Sousa, H.~Hansen and W.~M.~Alberico,
  Phys.\ Rev.\ D {\bf 79}, 116003 (2009).

\bibitem{Blaschke:2014zsa} 
  D.~Blaschke, A.~Dubinin and M.~Buballa,
  Phys.\ Rev.\ D {\bf 91}, 125040 (2015).
%
%
%

\end{thebibliography}
\end{document}